\documentclass[twocolumn,pra,aps,showpacs,10pt]{revtex4-1} 
\usepackage{amsmath,amssymb,graphicx}
\usepackage{hyperref}
\begin{document}
\title{Tunable fast and slow light in a hybrid optomechanical system}
\author{M. Javed Akram}\email{javed\_quaidian@yahoo.com; mjakram@qau.edu.pk}
\author{M. Miskeen Khan}
\author{Farhan Saif}
\affiliation{Department of Electronics, Quaid-i-Azam University, 45320 \ Islamabad, Pakistan.}
\begin{abstract}
We explain the probe field transmission spectrum under the influence of a strong pump field in a hybrid optomechanical system, composed of an optical cavity, a mechanical resonator, and a two-level atom. We show fast (superluminal) and slow (subluminal) light effects of the transmitted probe field in the hybrid system for suitable parametric regimes. For the experimental accessible domain, we find that the fast light effect obtained for the single optomechanical coupling can further be enhanced with the additional atom-field coupling in the hybrid system. Furthermore, we report the existence of a tunable switch from fast to slow light by adjusting the atomic detuning with the anti-Stokes and Stokes sidebands, respectively, as $\Delta_a=+\omega_m$ and $-\omega_m$. The reported characteristics are realizable in state-of-the-art laboratory experiments.
\end{abstract}
\maketitle 
\section{Introduction}
In recent years, tremendous progress in new methods for pulse propagation and properties of optical media have been reported in optomechanics. The rapidly emerging field has become a playground to study electromagnetically induced transparency (EIT) \cite{Harris}, inverse EIT \cite{Agarwal5}, stimulated Raman adiabatic passage \cite{Javed}, refractive index enhancement \cite{Scully,Yavuz}, Fano resonances \cite{Javed2,Ott,Agarwal2}, microscopy \cite{Javed4}, photonic crystals~\cite{Bigelow}, and nano-optic resonators \cite{Kippenberg}. In this paper, we explain the phenomena of parametrically controlled fast (superluminal) light and slow (subluminal) light in hybrid optomechanics, which is of considerable interest in view of its potential impact on present-day photonic technology \cite{Cheng,Mok,Kuzmich,Milonni,Gaeta}.

Various techniques have been developed to realize fast light and slow light in atomic vapors and solid-state materials. One application among these techniques is to control the group velocity $v_g$ of light pulses to make them propagate either very slow ($v_g < c$) or very fast ($v_g>c$ or $v_g$ is negative) \cite{Boyd1}. For instance, studies on slow light have made use of the technique of EIT in atomic vapors or Bose-Einstein condensate (BEC) \cite{Kasapi,Hakuta}. Astonishingly, Hau \textit{et al.} \cite{Hau} observed a dramatic reduction of the group velocity down to $17$ m$s^{-1}$ in an ultracold sodium vapour. Besides the slow light, superluminal (fast-light) phenomenon were observed in atomic cesium gas \cite{Wang} and in alexandrite crystal \cite{Boyd}. Safavi-Naeini {\it et al.} \cite{Painter} have observed the superluminal light with a $1.4$-$\mu s$ pulse advancement as well as tunable delay of the order $50$ $ns$ in a nanoscale optomechanical crystal device. Recently, Clark {\it et al}. \cite{Lett} reported the arrival time of the mutual information contained within the detection bandwidth in the fast and slow light media. More recently, Mirhosseini {\it et al.} \cite{Mirhosseini} have observed the dramatic enhancement in the fast light effect caused by electromagnetically induced absorption in warm rubidium vapor.

On the other hand, cavity optomechanics has surged swiftly through theoretical as well as experimental advancements over the last few years \cite{Meystre,AKM}. Combination of optomechanics with mechanical elements, like nano mechanical membranes \cite{Sankey,Regal,Javed3} and (ultra-cold) atoms \cite{Nori,Esslinger,Aspect}, leads to hybrid optomechanics \cite{Paternostro} which serves as a workhorse to study coherent dynamics in microscopic and macroscopic domains. Significant progress has been made in the investigation of various characteristics of optomechanics, such as quantum entanglement \cite{Ent}, quantum ground-state cooling \cite{Naik}, squeezing \cite{Regal,Regal2}, dynamical localization \cite{Kashif}, gravitational wave detection \cite{Caves}, EIT \cite{Agarwal,Javed2}, Fano resonances \cite{Agarwal2,Javed2} and classical dynamics \cite{Javed3,Chen}. Fast and slow light have also been observed in optomechanical systems \cite{Weis,Teufel,Stenner,Tarhan,Tarhan2,Zhu,Zhan,Jiang}, whose smaller dimensions and normal environmental conditions have paved the way towards real applications, such as telecommunication, interferometry, quantum-optomechanical memory and classical signal processing applications \cite{Painter2,Boyd3}.

In view of many potential applications of fast and slow light propagation, a question of interest is whether one can have a controlling parameter in a single set-up (experiment) for switching from superluminal to subluminal propagation or vice versa. However, previous studies show that single ended cavities allow only superluminal propagation \cite{Tarhan2}, whereas hybrid BEC \cite{Zhu}, quadratically coupled \cite{Zhan}, and even two-mode optomechanical systems \cite{Jiang} only allow slow light propagation. Furthermore, it is reported \cite{Tarhan} that, slow light in a double ended cavity occurs in the transmitted probe field, whereas the fast light effect takes place in the reflected field. In this paper, based on hybrid atom-cavity optomechanics, we report the following. \\
(i) There is a new and unique possibility to realize fast and slow light effects in the transmitted probe field by using a single set-up.
\\ (ii) We show that the additional atom-field coupling in the hybrid system, significantly enhances the superluminal behavior of the probe field obtained earlier for the single optomechanical coupling.
\\ (iii) In our experimentally feasible system, we provide a tunable switch that changes the propagation of light from superluminal to subluminal by adjusting the atomic detuning.
\\ Thus, in contrast to the earlier schemes, our proposal suggests a new and unique opportunity to realize fast and slow light effects in the transmitted probe field by using a single set-up.

The paper is organized as follows. In Sec.~\ref{sec2}, we present our hybrid system and give detailed analytical discussions; from the system Hamiltonian to the probe transmission spectrum via standard input-output theory. Section~\ref{sec3} is devoted to numerical results and discussions, where the occurrence of fast and slow light regimes have been explained. Finally, in Sec.~\ref{sec4}, we conclude our work.
\section{The Model System}\label{sec2}
\begin{figure}[ht]
\includegraphics[width=0.45\textwidth]{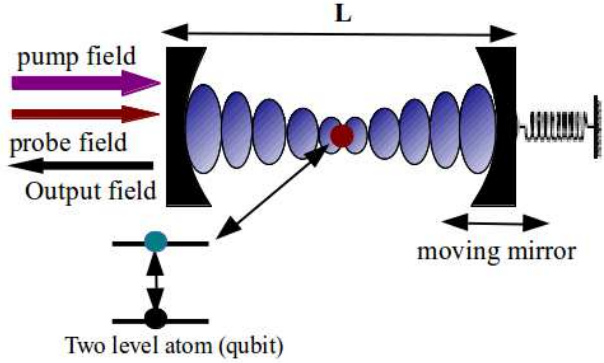}
\caption{(Color online) The schematic representation for the hybrid optomechanical system. A high $Q$ Fabry-P\'{e}rot cavity of length $L$ consists of a fixed mirror, a movable mirror and a two-level atom. The single-mode cavity field of frequency $\omega_c$ is coupled to both; the mechanical resonator as well as to the two-level atom. The cavity is simultaneously driven by a strong pump field of frequency $\omega_l$ and a weak probe field of frequency $\omega_p$.} \label{cavity}
\end{figure}

We consider a hybrid optomechanical system, composed of a high $Q$ Fabry-P\'{e}rot cavity of length $L$, a mechanical resonator, and a two-level atom, as shown in Fig.~\ref{cavity}. The single-mode cavity field is coupled to both; the mechanical resonator as well as the two-level atom. Hence, the system has the nonlinear optomechanical coupling and the Jaynes-Cummings (JC) (atom-field) coupling of cavity QED \cite{Haroche}.
The cavity is simultaneously driven by a strong pump field of frequency $\omega_l$ and a weak probe field of frequency $\omega_p$ along the cavity axis. Hence, under the rotating reference frame at the frequency $\omega_l$ of the strong driving field, the total Hamiltonian of the system can be written as \cite{Javed2}:
\begin{eqnarray}
H_T &=& [\dfrac{p^2}{2m} + \dfrac{1}{2} m \omega_m^2 q^2] + \hbar \Delta_c c^{\dagger}c +  \hbar \dfrac{\Delta_a}{2}\sigma_z - \hbar g_{mc}c^{\dagger}cq  \nonumber \\ &+&
\hbar g_{ac}(c^\dagger\sigma_- +c\sigma_+) + i\hbar E_l(c^\dagger - c) \nonumber \\ &+&
i\hbar(E_p e^{-i\delta t}c^\dagger - E_p^\ast e^{i\delta t}c)\label{Ham}.
\end{eqnarray}
Here, $\Delta_c=\omega_c-\omega_l$, $\Delta_a=\omega_a-\omega_l$, and $\delta=\omega_p-\omega_l$ are, respectively, detuning of the cavity field frequency, the atomic transition frequency, and the probe field frequency, with pump field frequency. In Eq.~(\ref{Ham}), the first term gives the free Hamiltonian of the moving mirror. Here, respectively, $q$ and $p$ represent the position and momentum operators with the vibration frequency $\omega_m$ and mass $m$. The second term describes the Hamiltonian of the cavity mode with creation (annihilation) operator $c^\dagger$ ($c$). The third term expresses the Hamiltonian of the two-level atomic system with transition frequency $\omega_a$. The next two terms describe the interaction Hamiltonian: the first term shows the interaction between the cavity and the oscillating mirror with optomechanical coupling strength $g_{mc}=\frac{\omega_c}{L}\sqrt{h/m\omega_m}$; the second term describes the atom-field interaction with (JC) coupling $g_{ac}$. The parameters $\sigma_+$ and $\sigma_-$ are the raising and lowering operators of the two-level atom with transition frequency $\omega_a$ and are expressed by $\sigma_\pm=\frac{\sigma_x \pm i\sigma_y}{2}$. Here $\sigma_{x,y,z}$ are the Pauli spin operators for the two-level atom. Finally, the last two terms in Eq.~(\ref{Ham}), correspond to the classical light fields (pump and probe lasers) with frequencies $\omega_l$ and $\omega_p$. Here, $E_l$ and $E_p$ are related to the laser power as $|E_l|=\sqrt{2\kappa P_{l}/ \hbar \omega_l}$ and $|E_p|=\sqrt{2\kappa P_{p}/ \hbar \omega_p}$, respectively, where $\kappa$ is the decay rate of the cavity.

After taking the corresponding dissipation and fluctuation terms into consideration, the Heisenberg-Langevin equations based on the Hamiltonian given in Eq.~(\ref{Ham}), are read as
\begin{eqnarray}
&&\dot{q}=\dfrac{p}{m}, \nonumber \\
&&\dot{p}=-m\omega_m^2 q - \gamma_m p +  g_{mc}c^\dagger c + \xi(t), \nonumber\\
&&\dot{c}=-(\kappa+i\Delta_c)c+ i g_{mc} c q-ig_{ac}\sigma_- +E_l \\ &&+E_p e^{-i\delta t} + \sqrt{2\kappa}c_{in}(t), \nonumber\\
&&\dot{\sigma}_-=-(\gamma_a+i\Delta_a)\sigma_- + ig_{ac} c \sigma_z+ \sqrt{2\gamma_a}a_{in}(t), \nonumber
\end{eqnarray}
where $\kappa$, $\gamma_a$, and  $\gamma_m$ denote the radiative decays associated with the cavity, the atom, and the mechanical mode, respectively. Here, we have introduced the input vacuum noises associated with the cavity field $c_{in}(t)$ and the atom $a_{in}(t)$ respectively, having zero mean values, i.e., $\langle c_{in}(t) \rangle = \langle a_{in}(t) \rangle=0$. Moreover, the input vacuum noises affecting the cavity field and the atom, obey the non-vanishing commutation relations \cite{Zoller}:
$\langle c_{in}(t) c_{in}^\dagger (t') \rangle = \delta(t-t')$ and $\langle a_{in}(t) a_{in}^\dagger (t') \rangle = \delta(t-t')$.
Furthermore, the Hermitian Brownian noise operator (thermal Langevin force) $\xi(t)$ with zero mean value $\langle \xi(t) \rangle =0$, satisfies the temperature-dependent correlation function \cite{Genes}: $\langle \xi(t) \xi^\dagger (t') \rangle = \int \omega e^{-i\omega (t-t')}N(\omega)d\omega$,
where, $N(\omega)=\frac{\gamma_m}{2\pi\omega_m} \left[ 1+coth \right( \frac{\hbar \omega}{2k_B T} \left) \right] $, $k_B$ is the Boltzmann constant, and $T$ is the temperature of the mechanical oscillator reservoir.

In order to explain the effect of the mechanical resonator and the atom on
the probe field transmission, we write the steady-state solutions of the corresponding operators and study the output spectrum. Using the mean field approximation \cite{Agarwal}, i.e., $\langle qc\rangle \simeq \langle q\rangle \langle c\rangle $, the mean value equations can be written as:
\begin{eqnarray}\label{mean}
&&\dfrac{d\langle p \rangle}{dt}=-m\omega_m^2 \langle q \rangle - \gamma_m \langle p \rangle + g_{mc}\langle c^\dagger \rangle \langle c \rangle, \nonumber\\
&&\dfrac{d\langle q \rangle}{dt}=\dfrac{\langle p \rangle}{m}, \nonumber\\
&&\dfrac{d\langle c \rangle}{dt}=-(\kappa+i\Delta_c) \langle c  \rangle +i g_{mc}  \langle c \rangle \langle q  \rangle -ig_{ac} \langle \sigma_- \rangle \\
&&+E_l + E_p e^{-i\delta t}, \nonumber\\
&&\dfrac{d\langle \sigma_- \rangle}{dt} =-(\gamma_a+i\Delta_a) \langle \sigma_- \rangle  + ig_{ac}  \langle c \rangle  \langle \sigma_z  \rangle. \nonumber
\end{eqnarray}
As we obtain the expectation values of the operators in the above set of equations, we drop the Hermitian Brownian noise and input vacuum  noise terms which are averaged to zero. In order to obtain the steady-state solutions of the above equations, we write the ansatz~\cite{Boyd2}
\begin{equation}\label{ansatz}
\langle h \rangle = h_s + h_- e^{-i\delta t} + h_+ e^{i\delta t},
\end{equation}
where, $h_s$ denotes any of the steady-state solutions $c_s$, $q_s$, and $a_s$. The above solution contains three components, which in the original frame, oscillate at frequencies $\omega_l$, $\omega_p$, and  $2\omega_l - \omega_p$, respectively. By substituting the ansatz into Eqs.~(\ref{mean}) and upon working to the lowest order in $E_p$, we obtain the following steady-state solutions:
\begin{eqnarray}\label{cs}
&&c_s=\dfrac{E_l}{\kappa + i\tilde{\Delta} - \frac{g_{ac}^2 \langle \sigma_z \rangle_{ss}}{\gamma_a+i\Delta_a}}, \\
&&c_-=\dfrac{E_p (A-B)}{B B' + (A-C)(A'+C)- (AB' +  A' B) + 2iC\tilde{\Delta}}, \nonumber  
\end{eqnarray}
where,
\begin{eqnarray}\label{set}
&&A =\kappa-i\Delta_c-i\tilde{\Delta} + \dfrac{i g_{mc}^2}{m\hbar(\omega_m^2-i\gamma_m \Delta-\Delta^2)}\vert c_s \vert^2, \nonumber\\
&&B =\dfrac{g_{ac}^2\langle \sigma_z \rangle_{ss}}{\gamma_a-i\Delta_a-i\Delta}, \nonumber\\
&&C =\dfrac{i g_{mc}^2}{m\hbar(\omega_m^2-i\gamma_m \Delta-\Delta^2)}\vert c_s \vert^2, \nonumber\\
&&\tilde{\Delta} =\Delta_c-\dfrac{g_{mc}^2}{m\hbar \omega_m^2}\vert c_s \vert^2.
\end{eqnarray}
Here, $A'=(A(-\Delta))^*$ and $B'=(B(-\Delta))^*$. The expressions given in Eq.~(\ref{cs}), lead us to study the probe transmission. Moreover, here we do not mention the expression for $c_+$ as this is associated with the four-wave mixing for the driving field and the weak probe field (which shall be discussed separately). Moreover, we are interested in the mean response of the system to the probe field, while there is the same detuning between the cavity field and the atomic system with respect to the pump field ($\Delta_{c}=\Delta_{a}$). Thus, without loss of generality, hereafter we assume that the atom stays in its excited state for $\Delta_{c}=\Delta_{a}$, which implies that $\omega_{c}=\omega_{a}$, that is, cavity field frequency is resonant with atomic transition frequency. This leads us to the conclusion that the atomic steady-state value is set as $\langle \sigma_z \rangle_{ss}=1$ \cite{Javed2,ss}. Furthermore, in the case of $\delta=\pm \omega_m$ or $\tilde{\Delta}$, the coupling between the moving mirror and the cavity field becomes stronger. To resolve simply, we consider that the system is in the sideband resolved limit, i.e., $\omega_m>>\kappa$ and $\delta \sim \omega_m$ \cite{Agarwal}.

In order to investigate the optical properties of the output field, we employ the standard input-output relation \cite{Weis}: $c_{out}(t) = c_{in}(t) - \sqrt{2\kappa}c(t)$, where $c_{in}$ and $c_{out}$ are the input and output operators, respectively. By using the above input-output relation and the ansatz given in Eq.~(\ref{ansatz}) for $\langle c(t) \rangle$, we can obtain the expectation value of the output field as,
\begin{eqnarray}
\langle c_{out}(t) \rangle&=&(E_l-\sqrt{2\kappa} c_s) + (E_p - \sqrt{2\kappa} c_-) e^{-i\delta t} \nonumber \\
&-& \sqrt{2\kappa} c_+ e^{i\delta t}. \label{out}
\end{eqnarray}
In analogy with Eq.~(\ref{ansatz}), the above solution of the output field contains three components. The first term corresponds to the output field at driving field frequency, $\omega_l$. The second term corresponds to the output field at probe frequency, $\omega_p$. The last term corresponds to the output field at frequency, $2\omega_l - \omega_p$, which corresponds to the Stokes field. It is generated via the nonlinear four-wave mixing process, in which two photons at frequency $\omega_l$ interact with a single photon at frequency $\omega_p$ to create a new photon at frequency $2\omega_l - \omega_p$ \cite{Agarwal4}. We note that, the second term on the right-hand side in expression (\ref{out}) corresponds to the output field at probe frequency $\omega_p$ with detuning $\delta$. Thus, the transmission of the probe field, which is the ratio of the returned probe field from the coupling system divided by the sent probe field \cite{Painter,Weis}, can be acquired as
\begin{equation}
T(\omega_p) = \frac{E_p - \sqrt{2\kappa} c_-}{E_{p}} = 1- \frac{\sqrt{2\kappa} c_-}{E_p}.
\end{equation}
For an optomechanical system, in the region of the narrow transparency window the rapid phase dispersion, viz., $\phi_t(\omega_p) = arg[T(\omega_p)]$, can cause the transmission group delay given by \cite{Weis},
\begin{equation} \label{phase}
\tau_g = \dfrac{d\phi_t(\omega_p)}{d\omega_p}=\dfrac{d\{arg[T(\omega_p)]\}}{d\omega_p}. 
\end{equation}
As the magnitude of group delay depends upon the rapid phase dispersion in the transmitted probe field, $\tau_g < 0$ and $\tau_g > 0$, corresponding to fast and slow light propagation, respectively. Therefore, the high phase dispersion is always advantageous, which allows for large change in the group delay, and leads to strongly altered group velocity \cite{Milonni,Clader}. In the following section, we show that hybrid optomechanics provide an effective way to achieve the high dispersion by controlling the atom-field coupling. Consequently, we anticipate that fast light obtained for the single ended cavities, can further be enhanced in the hybrid system. Moreover, we provide a tunable switch from fast to slow light by adjusting the atomic detuning.
\section{Results and Discussions}\label{sec3}
In order to quantify the superluminal (fast) and slow light effects, we consider experimentally realizable parametric values of the optomechanical system in our numerical simulations. We calculate the transmission spectrum, phase, and group delay of the probe field for the parameters \cite{Grolacher,AKM}: the optomechanical coupling $g_{mc}/2\pi=1.2$ MHz, $E_l/2\pi=2$ MHz, $\omega_m /2\pi=10$ MHz, $\kappa/2\pi=215$ kHz, $g_{ac}/2\pi=4$ MHz, $\Delta_c/2\pi=10$ MHz, $\Delta_a/2\pi=\pm 10$ MHz, $\gamma_a/2\pi=200$ kHz and $\gamma_m/2\pi=140$ Hz. Note that $\omega_m > \kappa$, therefore the system operates in the resolved-sideband regime, also termed the good-cavity limit \cite{Agarwal}. In the following, we explain the probe transmission, and observation of superluminal and slow light regimes in the hybrid system.

\subsection{Superluminal regime with single optomechanical coupling}
\begin{figure}[ht]
\centerline{\includegraphics[width=0.46\textwidth]{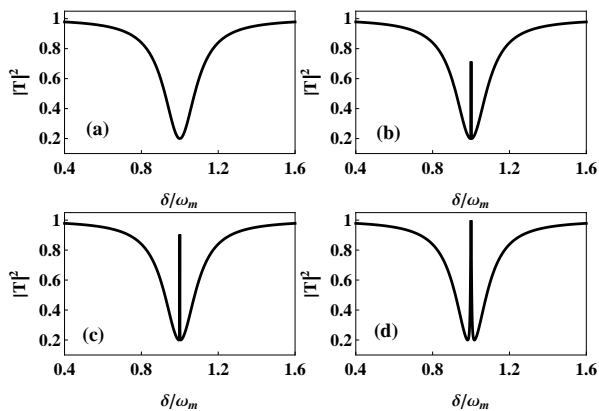}}
\caption{The transmission $|T|^2$ of the probe field as a function of normalized probe detuning is plotted for (a) to (d), $g_{mc}/2\pi=$0, 0.5, 0.8, and 1.2  MHz, respectively. The experimental values of the parameters are: $E_l/2\pi=2$ MHz, $\omega_m /2\pi=10$ MHz, $\Delta_c=\omega_m$, $\kappa=\omega_m/10$, and $\gamma_m/2\pi=140$ Hz \cite{Grolacher}. (Note that, $g_{ac}=0$ for this particular situation.)} \label{trans1}
\end{figure}

We consider the case when atom-field coupling is switched off, i.e. $g_{ac}=0$; this reduces the system to a single ended optomechanical system \cite{Tarhan2,Tarhan}. In Fig.~\ref{trans1}, we plot the transmission coefficient $|T|^2$  as a function of normalized probe detuning $\delta/\omega_m$, for different values of the optomechanical coupling $g_{mc}$. We see that, in the absence of the pump laser, i.e., $g_{mc}=0$, a standard Lorentzian curve appears in Fig.~\ref{trans1}(a) \cite{Weis,Javed2}. However, in the presence of a pump laser, the transmission spectrum of the probe field shows a prominent transparency window at the resonant region, viz., $\delta \sim \omega_m$, in Fig.~\ref{trans1}(b-d). The transparency window as appears in Fig.~\ref{trans1}, can be modulated effectively by the optomechanical coupling $g_{mc}$, or equivalently by the pump power \cite{Weis,Painter}. Such a physical phenomenon is very similar to EIT \cite{Agarwal,Javed2}.
\begin{figure}[t]
\includegraphics[width=0.45\textwidth]{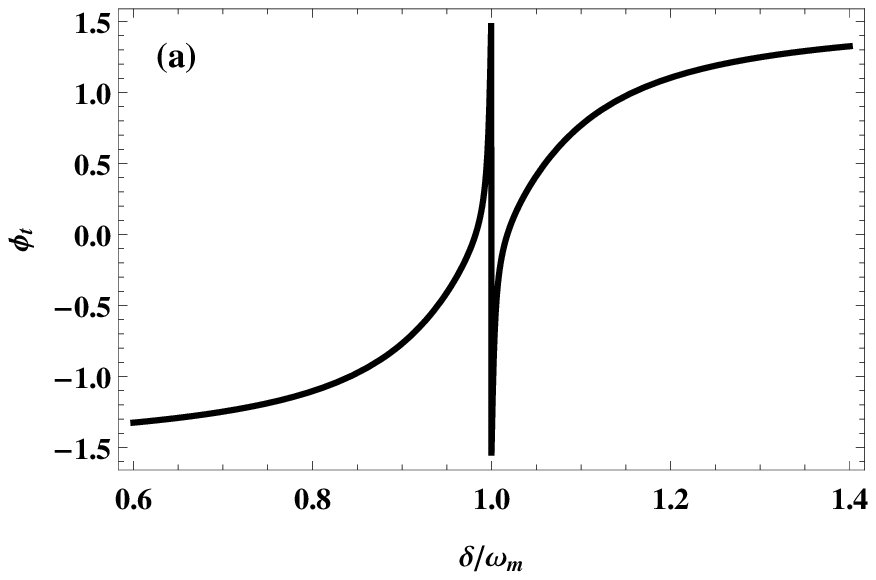}
\includegraphics[width=0.45\textwidth]{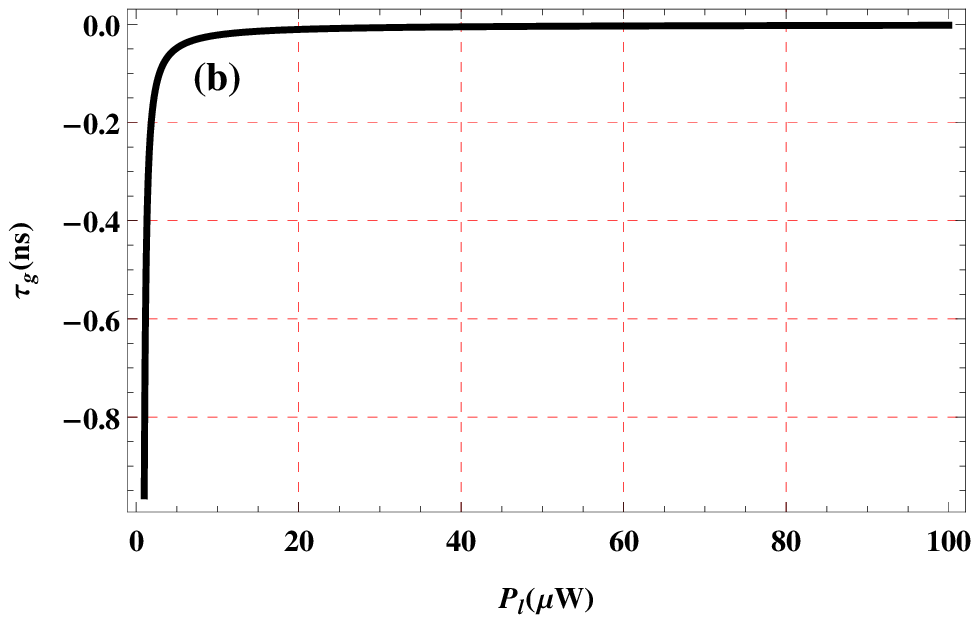}
\caption{(a) (Color online) The phase $\phi_t$, of the probe field as a function of normalized probe detuning, (b) Group delay $\tau_g$, as a function of the pump power ($P_l$), whereas $g_{ac}=0$. All parameters are the same as in Fig.~\ref{trans1}.} \label{delay1}
\end{figure}
The underlying physical mechanism for this phenomenon is as follows: the simultaneous presence of pump and probe fields generates a radiation pressure force at the beat frequency $\delta=\omega_p-\omega_l$, resonant with the mechanical frequency $\omega_m$. The frequency of the pump field $\omega_l$ is shifted to the anti-Stokes frequency $\omega_l + \omega_m$, which is degenerated with the probe field. Destructive interference between the anti-Stokes field and the probe field can suppress the build-up of an intracavity probe field and result in the narrow transparency window. The window becomes wider and more transparent when either the pump field power or the optomechanical coupling continuously increases \cite{Javed2}.

We plot the phase and group delay of the probe field in Fig.~\ref{delay1}. Within the transparency window, the phase of the transmitted probe beam suffers a sharp enhancement as shown in Fig.~\ref{delay1}(a), which indicates that the group velocity of the transmitted probe field can be strongly altered \cite{Clader,Milonni}. In Fig.~\ref{delay1}(b), we plot the group delay $\tau_g$ as a function of the pump power for $\delta=\omega_m$. We can see that by employing single coupling $g_{mc}$, the group delay is negative, which corresponds to the superluminal (fast-light) effect of the transmitted probe field. It is worth noticing that the phenomenon of superluminal effect can be counterintuitive owing to the presence of phenomena for which it is possible that the peak of the output pulse may exit the optical material before it passes through the entrance face \cite{Garrett,Clader}. Its consistency with the principle has been verified experimentally by Stenner {\it et al.} \cite{Stenner}. Thus, the effect of single optomechanical coupling leads to the superluminal propagation in the probe field transmission as noted earlier \cite{Tarhan2,Tarhan}.

\subsection{Superluminal regime in hybrid optomechanics}
\begin{figure}[ht]
\centering{\includegraphics[width=0.45\textwidth]{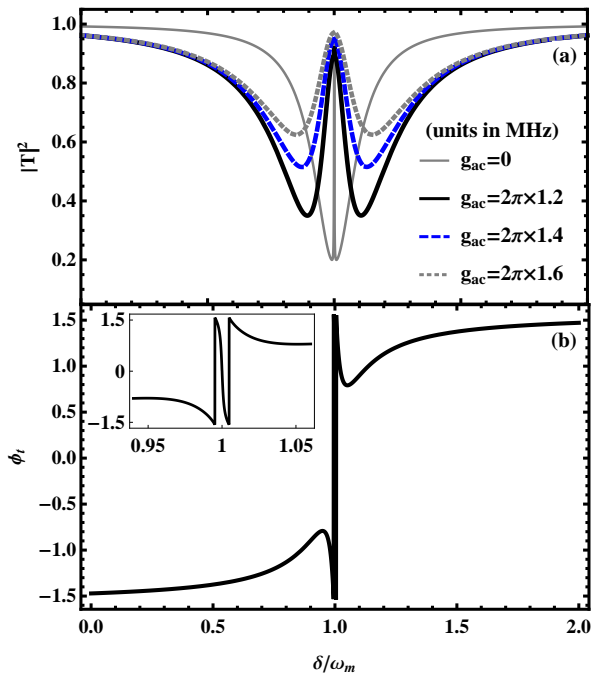}}
\caption{(Color online) (a) The transmission $|T|^2$ of the probe field as a function of normalized probe detuning is shown, involving both coupling parameters of the hybrid system. The experimental parameters of the atomic system are: $g_{ac}=0$ for the thin gray curve,  and for the rest $g_{ac}/2\pi=$1.2, 1.4, 1.6 MHz (from bottom to top) with $\Delta_a=\omega_m$ and $\gamma_a/2\pi=200$ kHz. (b) Phase $\phi_t$ of the probe field as a function of normalized probe detuning is shown for $g_{ac}/2\pi=1.2$ MHz. The inset shows the magnified phase $\phi_t$ for the same parametric values. Rest of the parameters are the same as in Fig.~\ref{trans1}.} \label{trans2}
\end{figure}

Here, we analyze how the addition of the two-level atom in the optical cavity can influence the transmission of the probe field. In Fig.~\ref{trans2}, we show the transmission and the phase of the probe field as a function of the normalized probe detuning $\delta/\omega_m$ for $P_l =6$ $\mu$W and $\Delta_a=\omega_m$. Fig.~\ref{trans2}(a) plots the transmission spectrum for different values of JC coupling $g_{ac}$. We see that a narrow transparency window appears for $g_{ac}=0$, (thin gray curve). However, different from the output field for the single coupling as in Fig.~\ref{trans1}, the characteristics of the probe field changes in the presence of JC coupling in the present case.
From Fig.~\ref{trans2}(a), one can see that the transmission window becomes wider and wider by continuously increasing the atom-field interaction. Hence, one can estimate the magnitude of the atom-field interaction by measuring the width of the window. It is also worth noticing that, different from the single coupling case, when a two-level atom is coupled to the opto-mechanical resonator, the atomic state splits into two states making the dressed states by the single-photon state and the two-level system for $\omega_a = \omega_c$ \cite{Haroche}. This splitting of the atomic state significantly affects the probe field transmission. Moreover, due to the addition of the atom in the system ($g_{ac}\neq 0$), Eqs.~(\ref{cs}) and (\ref{set}) reveal that the effective loss of the cavity and the effective detunning $\tilde{\Delta}$ will decrease, which is equal to enhancing the radiation pressure \cite{Nori2}. Thus, the existence of atom-field coupling, together with $g_{mc}$, significantly affects the probe transmission and may lead to the high phase dispersion that can strongly alter the group velocity of the transmitted probe field.

In Fig.~\ref{trans2}(b), we plot the phase of the transmitted probe field in the presence of the two-level atom in the optical cavity. Figure~\ref{trans2}(b) and the inset show that, a relatively high and rapid phase change occurs in the resonant interval of the transparency window ($\delta\sim \omega_m$) as compared with the case of single coupling shown in Fig.~\ref{delay1}(b). The phase dispersion curve for $\Delta_a=+\omega_m$ represents the anomalous dispersion as shown in Fig.~\ref{trans2}(b). Moreover, in the context of fast and slow light, the high phase dispersion is always advantageous as it greatly alters the group index of the medium, through which a probe pulse will propagate with strongly altered group velocity \cite{Clader}. Unlike single ended cavities, here we show that the high dispersion can be achieved by simply adjusting the JC coupling in the hybrid system, which may result in an enhancement of superluminal behavior. 
\begin{figure}[t]
\includegraphics[width=0.45\textwidth]{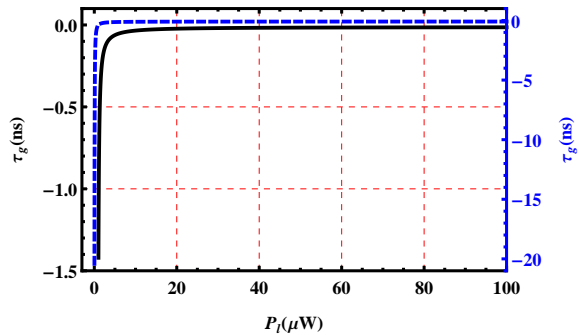}
\caption{(Color online) Group delay ($\tau_g$) as a function of the pump power for $\delta=\omega_m$ and $\Delta_a=\omega_m$, is shown for different values of the atom-field coupling $g_{ac}$. The black-solid and blue-dashed curves correspond to the pulse advancement for $g_{ac}/2\pi=4$ and 8 MHz, respectively. All other parameters are the same as in Fig.~\ref{trans2}.} \label{delay2}
\end{figure}
\begin{figure*}[t]
\includegraphics[width=0.45\textwidth]{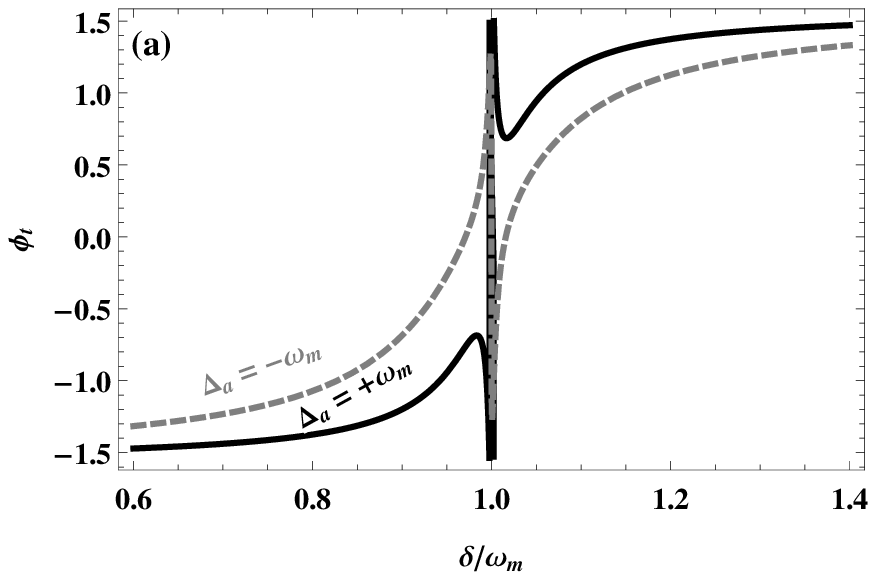}
\includegraphics[width=0.45\textwidth]{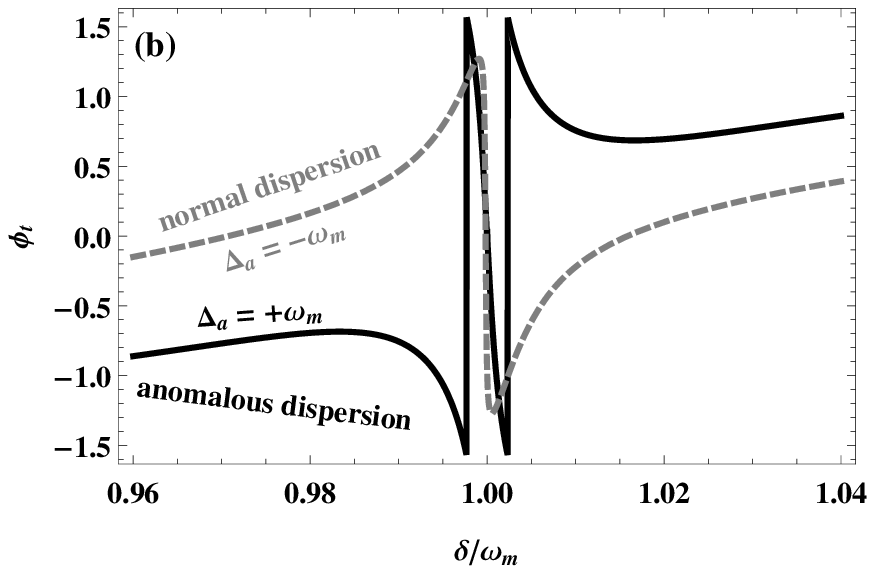}
\caption{(a) The phase of the probe field, $\phi_t$, versus the normalized probe detuning $\delta/\omega_m$, for $\Delta_a=\omega_m$ (black solid curve) and $\Delta_a=-\omega_m$ (gray dashed curve) for $g_{ac}/2\pi=1.6$ MHz and $g_{mc}/2\pi=0.1$ MHz. (b) We show the magnified dispersion curves for the same parametric values showing the dispersion dynamics in the resonant region. The switching of atomic detuning changes the dispersion curve from anomalous for $\Delta_a=+\omega_m$ to normal for $\Delta_a=-\omega_m$. The rest of the parameters are the same as in Fig.~\ref{trans2}.} \label{delay3}
\end{figure*}

In Fig.~\ref{delay2}, the group delay $\tau_g$ is plotted as a function of the pump power with $\delta = \omega_m$ and $\Delta_a = \omega_m$, for different values of JC coupling. For the experimental accessible parameters, we obtain the pulse advancement of the order $1.2$ ns for $g_{ac}=2\pi \times 4$ MHz (black curve), and $20$ ns for $g_{ac}=2\pi \times 8$ MHz (blue dashed curve). Note that, as compared to the previous case of single coupling in Fig.~\ref{delay1}(b), here we achieve the pulse advancement of the order $20$ ns, significantly larger in magnitude. Thus, we infer that the pulse advancement increases by continuously increasing the JC coupling under a constant driving field in the hybrid system. Hence, by employing both coupling parameters simultaneously, we achieve the high phase dispersion which leads to the higher pulse advancement than single ended cavities.

\subsection{Subluminal regime in hybrid optomechanics}
In the context of fast and slow light, a question of interest is whether one can have a controlling parameter in a single set-up for switching from superluminal to subluminal propagation or vice versa. In 2001, Agarwal {\it et al.}~\cite{Agarwal6}, and thereafter in 2004, Sahrai {\it et al.} \cite{Zubairy2}, proposed a switching mechanism from subluminal to superluminal propagation based on a $\Lambda$-type system and four-level atomic system, respectively. Analogous to the multi-level atomic systems, single ended optomechanical systems share the properties of three-level atomic system, and therefore, lead to the occurrence of a single EIT window \cite{Agarwal,Javed2}. The occurrence of a double EIT window, for example, or the switching mechanism from the superluminal to subluminal regime, requires additional interfering pathways \cite{Agarwal6,Zubairy2}, and the hybrid optomechanical systems spectacularly meet this requirement, as they can share the properties of four (or more) level atomic systems \cite{Javed2, Wang2}.

In our scheme, change in the atomic detuning from $\Delta_a=\omega_m$ to $-\omega_m$, acts as a tunable switch from superluminal to subluminal light propagation in the hybrid system. The atomic detuning $\Delta_a=- \omega_m$ means that the two-level atom is resonant with the Stokes sideband.

In Fig.~\ref{delay3}(a), we plot the phase $\phi_t$ versus normalized probe detuning $\delta/\omega_m$ for $\Delta_a = \pm \omega_m$. It can be seen that the phase of the probe field changes suddenly with a steeper dispersion as we switch the atomic detuning from $\Delta_a=\omega_m$ (solid black curve) to $\Delta_a=-\omega_m$ (dashed gray curve). Figure~\ref{delay3}(b) shows the magnified dispersion curves for the same parametric values, showing the dispersion dynamics in the resonant region. We see that the slope of the dispersion curve goes from negative to positive as we adjust the atomic detuning with the Stokes sideband. The switching of atomic detuning changes the dispersion curve from anomalous for $\Delta_a=+\omega_m$ to normal for $\Delta_a=-\omega_m$. This reflects that the propagation dynamics of the probe field alters from superluminal to subluminal propagation.

\begin{figure}[t]
\includegraphics[width=0.45\textwidth]{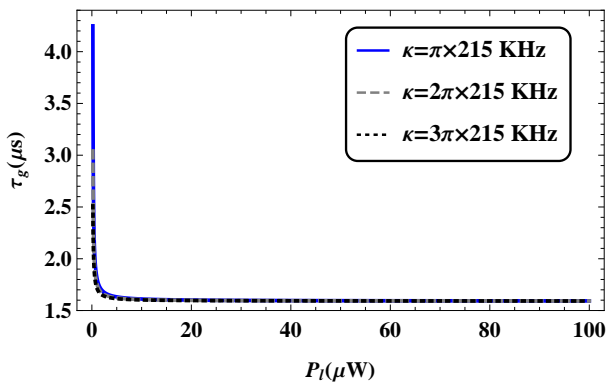}
\caption{(Color online) Group delay ($\tau_g$) as a function of the pump power ($P_l$) for the case of $\Delta_a=-\omega_m$, indicating the slow light effect of the transmitted probe field for the different values of cavity decay rate $\kappa=\pi \times 215, 2\pi \times 215, 3\pi \times 215$ kHz, for blue solid, gray dashed and black dotted curves, respectively. The rest of the parameters are the same as in Fig.~\ref{trans2}.} \label{f7}
\end{figure}

The group delay, $\tau_g$, versus the pump power is plotted in Fig.~\ref{f7} for $\Delta_a=-\omega_m$, against various values of cavity decay rate $\kappa$. The group delay is positive, which confirms that the characteristics of the transmitted probe field changes from superluminal to subluminal light as we turn atomic detuning as $\Delta_a=-\omega_m$. The magnitude of group delay increases with decrease in $\kappa$, as shown in Fig.~\ref{delay3}(b), which shows that the longer the lifetime of the resonator the more obvious we see the slow light effect.

In this particular regime, i.e., $\Delta_a =-\omega_m$, Genes \textit{et al.}~\cite{Genes} has found the largest stationary entanglement between atoms and mirror. It has been found that one has a sort of entanglement sharing: due to the presence of the atom(s), the initial cavity-mirror entanglement partially redistributes to the atom-mirror and atom-cavity subsystems and this effect is predominant when the atoms are resonant with the Stokes sideband. Here, we achieve a tunable switch from superluminal to slow light with $\Delta_a= \pm \omega_m$. Thus, consistency of the switching mechanism from superluminal to slow light regime ($\Delta_a =-\omega_m$) is in agreement with the realistic experimental conditions \cite{Genes}. In addition, the orders of pulse advancement and delay, i.e., $\tau_g=-20$ ns and $\tau_g=4.2$ $\mu$s, show a very good agreement with the experimental realization of fast and slow light in Refs.~\cite{Stenner,Painter}, which can be further enhanced with the longer lifetime of the resonator.

Hence, observation of superluminal light and its enhancement, and existence of a tunable switch from superluminal to slow light, make our model more practical and advantageous over previous schemes, such as, the single ended cavities \cite{Tarhan2}, coupled BEC-cavity systems \cite{Zhu}, quadratically coupled systems \cite{Zhan}, and hybrid optomechanics with two optical modes \cite{Jiang}. Moreover, we explain the fast and slow light effects in the probe field transmission. However, previously it was reported \cite{Tarhan} that, slow light in a double ended cavity occurs in the transmitted probe field, whereas the fast light effect takes place in the reflected field. The recent progress in optomechanics, makes it possible to realize our reported characteristics in state-of-the-art laboratory experiments.
\section{Conclusion}\label{sec4}
In conclusions, we have explained tunable fast and slow light effects of a transmitted probe field in a hybrid optomechanical system for experimentally realizable parametric values. Based on standard input-output theory, the full analytical model is presented to investigate the transmission, phase, and group delay of the probe field. It is shown that the addition of a two-level atom in the system, not only affects the transmission of the probe field, but also yields the high phase dispersion, which makes it possible to realize the enhancement of superluminal light in the hybrid system. In addition, a tunable switch from superluminal to slow light is achievable in our model by simply adjusting the atomic detuning as $\Delta_a=\pm \omega_m$, which makes our scheme thereby more advantageous and more practical over earlier schemes. Fast and slow light effects have potential impact on the present-day photonic technology and have paved the way towards many applications including quantum information processing, integrated quantum optomechanical memory, classical signal processing, real quality imaging, cloaking devices, higher detection efficiency of x rays, optical buffering, delay lines, telecommunication and interferometry \cite{Glasser,Milonni,Painter,Heeg}.
\section*{Acknowledgments}
We submit our thanks to K. Hakuta and G. Alber for valuable suggestions in our manuscript. We also acknowledge Higher Education Commission, Pakistan and Quaid-i-Azam University for financial support through Grants No. \#HEC/20-1374, and No. QAU-URF2014. 


\begin{thebibliography}{99}
\bibitem{ss} Note that, $\langle \sigma_z \rangle_{ss}$ is a number which can rescale the value of $g_{ac}$, see for instance Eqs. (\ref{cs},\ref{set}). However, it does not affect the occurrence of rapid phase change or the tunable switching mechanism until it is non-zero. The same assumption can also be reconciled by pumping the cavity with an additional field of frequency near resonance with the atomic transition frequency.
\end{thebibliography}
\end{document}